\documentclass{ws-ijmpa}
\usepackage[super,compress]{cite}
\usepackage{graphicx}
\begin{document}
\markboth{M. J. Guzm\'an \& R. Ferraro}{Degrees of freedom and Hamiltonian formalism for $f(T)$ gravity}

%
\catchline{}{}{}{}{}
%

\title{Degrees of freedom and Hamiltonian formalism for $f(T)$ gravity}

\author{Mar\'ia Jos\'e Guzm\'an}
\address{Departamento de F\'isica y Astronom\'ia, Facultad de Ciencias, Universidad de La Serena, Av.~Juan Cisternas 1200, La Serena 1720236, Chile\\
maria.j.guzman.m@gmail.com}
\author{Rafael Ferraro}
\address{Instituto de Astronom\'ia y F\'isica del Espacio (IAFE), CONICET, Universidad de Buenos Aires, Casilla de Correo 67, Sucursal 28, Buenos Aires 1428, Argentina\\
Departamento de F\'isica, Facultad de Ciencias Exactas y Naturales, Universidad de Buenos Aires, Ciudad Universitaria, Pabell\'on I, Buenos Aires 1428, Argentina\\
ferraro@iafe.uba.ar}
\maketitle

\begin{history}
\received{Day Month Year}
\revised{Day Month Year}
\end{history}

\begin{abstract}
The existence of an extra degree of freedom (d.o.f.) in $f(T)$ gravity has been recently proved by means of the Dirac formalism for constrained Hamiltonian systems. We will show a toy model displaying the essential feature of $f(T)$ gravity, which is the pseudo-invariance of $T$ under a local symmetry, to understand the nature of the extra d.o.f.

\end{abstract}

\ccode{PACS numbers: 04.50.Kd, 11.10.Ef}


\section{$f(T)$ Gravity}

The teleparallel equivalent of general relativity (TEGR) is a reformulation of general relativity (GR) in terms of a field of tetrads. It encompasses the vector basis $\mathbf{e}_a = e_a^{\mu} \partial_{\mu}$ and its co-basis $\mathbf{E}^a = E^a_{\mu} dx^{\mu}$, which are mutually dual: $E^a_{\mu} e_b^{\mu} = \delta^a_b$. Tetrads are related to the spacetime metric through the orthonormality condition
\begin{equation}
\eta_{ab} = g_{\mu\nu} e^{\mu}_a e^{\nu}_b~, \ \ \ \ \ g_{\mu\nu} = \eta_{ab} E^a_{\mu} E^b_{\nu}~.
\end{equation}
The spacetime underlying TEGR is endowed with a curvatureless, metric-compatible spin connection. Usually the Weitzenb\"{o}ck connection $\omega^a_{\ b \mu}=0$ is chosen, which in coordinate bases means $\Gamma^{\rho}_{\ \mu\nu} = e^{\rho}_a \partial_{\mu} E^a_{\nu}$. TEGR Lagrangian is built from the torsion $T^{\rho}_{\ \mu\nu} = e^{\rho}_a ( \partial_{\mu} E^a_{\nu} - \partial_{\nu} E^a_{\mu} )$ through the torsion scalar $T$ defined as \cite{Aldrovandi2013}
\begin{equation}
T = -\dfrac{1}{4}T_{\rho\mu\nu} T^{\rho\mu\nu} -\dfrac{1}{2} T_{\rho\mu\nu} T^{\mu\rho\nu} + T^{\rho}_{\ \mu\rho} T^{\sigma\mu}_{\ \ \sigma}~.
\label{tscalar}
\end{equation}
TEGR Lagrangian $L=ET$ ($E$ stands for $ \text{det}(E^a_{\mu})=|g|^{1/2}$) and GR Lagrangian $L=-ER$ ($R$ being the Levi-Civita scalar curvature) are dynamically equivalent since they differ in a boundary term: $E (R+T) = \partial_{\mu}(E T^{\nu \ \mu}_{\ \nu} )$. So, both TEGR and GR govern the same d.o.f., which are associated with the metric tensor. The metric tensor is invariant under local Lorentz transformations of the tetrad, $\mathbf{E}^a \rightarrow \mathbf{E}^{a^{\prime}} = \Lambda^{a^{\prime} }_{\ a}(x) \mathbf{E}^a$, which is thus a gauge symmetry of TEGR. The TEGR Lagrangian is used as a starting point to describe generalizations to GR inspired in $f(R)$ theories; the so called $f(T)$ gravity is governed by the action \cite{Ferraro:2006jd}
\begin{equation}
S = \dfrac{1}{2\kappa} \int d^4 x\ E\ f(T).\label{action}
\end{equation}

\section{A Toy Model with Rotational Pseudo-Invariance}
TEGR Lagrangian $L=ET$ is not gauge invariant but pseudo-invariant, because $T^{\nu \ \mu}_{\ \nu}$ in the above mentioned boundary term is not invariant under local Lorentz transformations of the tetrad. Therefore, a general function $f$ will not allow the boundary term to be integrated out in the $f(T)$ action \eqref{action}; as a consequence, the theory will suffer a partial loss of the local Lorentz symmetry;\cite{Ferraro:2006jd} so an extra d.o.f. not related to the metric could appear. We will analyze this issue by resorting to a simple toy model with rotational pseudo-invariance (a similar one was introduced in a previous work \cite{Ferraro:2018axk}, but the boundary term was simpler). Let be the Lagrangian
\begin{equation}
L = 2\left(\dfrac{d}{dt} \sqrt{z\overline{z}} \right)^2 - U(z\overline{z}) + \dot{z} \dfrac{\partial}{\partial z} g(z,\overline{z})+\dot{\overline{z}} \dfrac{\partial}{\partial \overline{z}} g(z,\overline{z})~.
\label{Ltoymod}
\end{equation}
The two first terms are invariant under local rotations $z \rightarrow e^{i\alpha(t)}z$. The rest of $L$ is a total derivative; it does not take part in the dynamics but can be affected by the local rotation. So, the Lagrangian $L$ is just {\it pseudo-invariant} under a local rotation. As any gauge invariance the local pseudo-invariance implies the existence of constraints among the canonical momenta; a unique primary constraint is obtained in this case:
\begin{equation}
G^{(1)} \equiv z\left(p_z - \dfrac{\partial g}{\partial z} \right) - \overline{z}\left( p_{\overline{z}} - \dfrac{\partial g}{\partial\overline{z} } \right) \approx 0.\label{G1}
\end{equation}
$G^{(1)}$ is an angular momentum; it generates rotations. In fact, it is $\{ G^{(1)},z\overline{z} \} = 0$, which means that the dynamical variable $|z|$ is gauge invariant. As can be seen, the angular momentum not only is conserved in this case; since the symmetry is local (time-dependent), the conserved value is constrained to be zero.

Primary constraints have to be consistent with the evolution, as controlled by the primary Hamiltonian $H_p = H + u(t) G^{(1)}$. In the case \eqref{Ltoymod}-\eqref{G1} it results that the consistency is fulfilled without specifying the Lagrange multiplier $u(t)$. Thus, the evolution of any variable that does not commute with $G^{(1)}$ is affected by an undetermined function $u(t)$; this is the case of the phase of $z$, which become a ``pure gauge" variable, but not the case of $|z|$, which is a genuine d.o.f. or observable. $G^{(1)}$ is called {\it first-class}, since it commutes with all the constraints (it is the only constraint in this example). As it is well known, each first class constraint removes one d.o.f. from a Hamiltonian constrained system. In this toy model, one d.o.f. is removed from the pair $(z,\overline{z})$, showing that $|z|$ is the only d.o.f. of the theory.

\section{Modified toy model}
We will {\it deform} the toy model of the previous section by introducing the Lagrangian $f(L)$. Let us show that this can be done by means of the Lagrangian 
\begin{equation}
\mathcal{L} = \phi L - V(\phi),
\label{modLtm}
\end{equation}
where $\phi$ is an auxiliary canonical variable. Equation \eqref{modLtm} resembles the Jordan-frame representation of $f(R)$ gravity. From $\mathcal{L}$ one gets the equation of motion for $\phi$: $L=V^{\prime}(\phi)$. Thus, $\mathcal{L}$ is (on-shell) equal to the Legendre transform of $V(\phi)$; therefore it depends only on $L$, i.e. $\mathcal{L}=f(L)$ (from the inverse Legendre transform we also know that $\phi=f^{\prime}(L)$). Thus the Lagrangian $\mathcal{L}$ is dynamically equivalent to a $f(L)$ theory. As expected for a $f(L)$ theory, $\mathcal{L}$ is not pseudo-invariant under local rotations. This is because the total derivative coming with $L$ is now multiplied by $\phi$ in \eqref{modLtm}. We will present the main outcomes of the Hamiltonian formalism for this $f(L)$ model and see the implicancies of the lost pseudo-invariance. 

By computing the canonical momenta for  $\mathcal{L}$ one gets two primary constraints: the angular momentum and the momentum conjugated to $\phi$,
\begin{equation}
G^{(1)}  = z \left( p_z - \phi \dfrac{\partial g}{\partial z} \right) - \overline{z} \left( p_{\overline{z}} - \phi \dfrac{\partial g}{\partial \overline{z}} \right) \approx 0, \ \ \ \ \ \ \ \ \ G^{(1)}_{\pi} = \pi = \dfrac{\partial \mathcal{L} }{\partial \dot{\phi}} \approx 0.
\end{equation}
The Poisson bracket between the constraints is
\begin{equation}
\{G^{(1)}, G^{(1)}_{\pi} \} = -z \dfrac{\partial g}{\partial z} + \overline{z} \dfrac{\partial g}{\partial \overline{z}},
\label{pbmodtoy}
\end{equation}
which depends on the function $g(z,\overline{z})$ appearing in the boundary term of $L$. Depending on $g$, the Poisson bracket could be zero or not, which would drastically affect the counting of d.o.f. So, we will separate two cases:
\begin{itemize}
\item \textbf{Case (i): $g(z,\overline{z}) \neq v(z\overline{z})$.} In this case it is $\{G^{(1)}, G^{(1)}_{\pi} \}\not\approx 0$, so the constraints are {\it second class}. The consistency is guaranteed by choosing the Lagrange multipliers $u^{\pi}(t)$ and $u(t)$ associated with $G_{\pi}$ and $G^{(1)}$, respectively. In particular, it results $u^{\pi}=0$ which implies that $\phi$ does not evolve but is a constant. The constancy of $\phi$ also implies that $|z|$  evolves like in the undeformed theory governed by $L$. But now the evolution of the phase of $z$ is determined too, because the Lagrange multiplier $u(t)$ is no longer left free. Since the evolution is already consistent at this step, then the algorithm is over. The counting of d.o.f. goes like this: from the set of three canonical variables $(\phi,z,\overline{z})$, just {\it one} d.o.f. is removed due to the appearance of {\it one pair} of second class constraints. We are left with two d.o.f., which can be represented by the variables $(z,\overline{z})$. The Lagrangian $f(L)$ determines not only the modulus of $z$ but its phase as well.

\item \textbf{Case (ii): $g(z,\overline{z})=v(z\overline{z})$.} In this case it is $\{G^{(1)}, G^{(1)}_{\pi} \}= 0$. This case is trivial because if $g(z,\overline{z})=v(z\overline{z})$ the entire Lagrangian $L$ will depend exclusively on $|z|$, so being locally invariant. Thus we do not expect an extra d.o.f. in the deformed $f(L)$ theory. So, let us check that Dirac's algorithm yields the right answer. The consistency of the constraints with the evolution leads to a new {\it secondary} constraint $G^{(2)} = L - V^{\prime}(\phi)\approx 0$. Since $\{G^{(1)}, G^{(2)} \} = 0$, and $\{G^{(1)}_{\pi}, G^{(2)} \} = V^{\prime \prime}(\phi)$, then $G^{(1)}$ is first-class, while $G^{(1)}_{\pi}$, $G^{(2)}$ are second-class. The Lagrange multiplier $u^{\pi}(t)$ is fixed by the consistency equations. Instead $u(t)$ (associated with  $G^{(1)}$ in $H_p$) is not fixed by the algorithm, so meaning that the variables that are sensitive to rotations, like the phase of $z$, will remain as pure gauge variables. The counting of d.o.f. goes like this: from the three canonical variables $(\phi,z,\overline{z})$ we remove two d.o.f., one coming from $G^{(1)}$ being first-class, and the other one because the pair $G^{(1)}_{\pi}$, $G^{(2)}$ is second-class, leaving us with the genuine d.o.f. $|z|$. Remarkably,  $u^{\pi}(t)$ results in a non zero function; therefore $\phi$ is not a constant and affects the evolution of $|z|$, that departs from the evolution it had in the original undeformed theory $L$. 
\end{itemize}

\subsection{Conclusions}

In principle $f(T)$ gravity is case-(i), since TEGR Lagrangian is pseudo-invariant under local Lorentz transformations 
of the tetrad. This means that $f(T)$ gravity entails an extra d.o.f. associated with the orientation of the tetrad. However we could wonder whether $f(T)$ gravity can be case-(ii) on-shell. This is an interesting point because, even though $f(T)$ gravity is case-(i), there could exist particular solutions to the equations of motion making zero the value of the Poisson bracket \eqref{pbmodtoy}. For such solutions, $\phi$ (and so $T$ too) would be an evolving field, and no extra d.o.f. would manifest. Remarkably, flat FRW spacetime seems to be a good arena to test this conjecture, because it contains both solutions with $T$ equal to a constant\cite{Bejarano:2017akj} and $T=-6 H^2(t)$ an evolving function\cite{Ferraro:2006jd}.

\section*{Acknowledgments}
M.J.G. has been funded by CONICYT-FONDECYT Postdoctoral grant No. 3190531. R.F. has been funded by CONICET and Universidad de Buenos Aires. R.F. is a member of Carrera del Investigador Cient\'ifico (CONICET, Argentina).


\end{document}